\documentclass[11pt]{article}
\usepackage[dvipdfmx]{color}
\usepackage{graphicx}
\usepackage{xcolor}
  \usepackage{amsthm,amsfonts}
  \usepackage{amsmath}
\usepackage{mathabx}
\usepackage{slashed}
\usepackage{ulem}
\usepackage{hyperref}
\usepackage{cleveref}
\usepackage{cite}

\newcommand{\bea}   {\begin{eqnarray}}
\newcommand{\eea}   {\end{eqnarray}}
\def\zzg{${\mathbb Z}_2\times{\mathbb Z}_2$-graded }

\begin{document}
\renewcommand{\thefootnote}{\fnsymbol{footnote}}

\thispagestyle{empty}

\title{First quantization of braided Majorana fermions}
\author{ Francesco Toppan\thanks{{E-mail: {\it toppan@cbpf.br}}}
\\
\\
}
\maketitle

{\centerline{
{\it CBPF, Rua Dr. Xavier Sigaud 150, Urca,}}\centerline{\it{
cep 22290-180, Rio de Janeiro (RJ), Brazil.}}
~\\
\maketitle

\begin{abstract}
A ${\mathbb Z}_2$-graded qubit represents an even (bosonic) ``vacuum state" and an odd, excited, Majorana fermion state. The multiparticle sectors of $N$, braided, indistinguishable Majorana fermions are constructed via  first quantization. The framework is that of a graded Hopf algebra endowed with a braided tensor product.  The Hopf algebra is ${\cal U}({\mathfrak {gl}}(1|1))$, the Universal Enveloping Algebra of the ${\mathfrak{gl}}(1|1)$ superalgebra. A $4\times 4$ braiding matrix $B_t$ defines the braided tensor product. $B_t$, which is related to the $R$-matrix of the Alexander-Conway polynomial, depends on the braiding parameter $t$ belonging to the punctured plane  ($t\in {\mathbb C}^\ast$); the ordinary antisymmetry property of fermions is recovered for $t=1$. \\
For each $N$, the graded dimension $m|n$ of the graded multiparticle Hilbert space is computed. Besides the generic case, truncations occur when $t$ coincides with certain roots of unity which appear as solutions of an ordered set of polynomial equations.   The roots of unity are organized into levels which specify
the maximal number of allowed braided Majorana fermions in a multiparticle sector. \\
By taking into account that the even/odd sectors in a ${\mathbb Z}_2$-graded Hilbert space are superselected, a nontrivial braiding with $t\neq 1$ is essential to produce a nontrivial Hilbert space described by qubits, qutrits, etc., 
since at $t=1$ the $N$-particle vacuum and the antisymmetrized excited state  encode the same information carried by a classical $1$-bit.  
\end{abstract}
\vfill
\rightline{CBPF-NF-002/22}
\newpage

\section{Introduction}

Majorana fermions and their braiding properties started being intensively investigated since Kitaev's proposal presented in \cite{kit}. In that work it was suggested to use them for encoding logical operations of a topological quantum computer which offers protection from decoherence (see also \cite{brki}). Several aspects of this proposal and the knot logic underlying it are discussed in \cite{kau}.\par
In this paper I discuss a framework, which can be called the ``first quantization of braided Majorana fermions", to derive their quantum properties. It is a new application (based on a graded Hopf algebra and with braiding properties being encoded in a braided tensor product) of the formalism introduced by Majid in \cite{maj}. This scheme substantially differs from other approaches to braided Majorana fermions as those discussed in \cite{kalo} (see also references therein) and  \cite{geli,gyxz} (in these two papers the focus is on the Yang-Baxter equation). The main differences are summarized as follows: a multiparticle quantum mechanics is derived and, for each $N$-particle sector, the graded dimension $m|n$ of the graded Hilbert space is computed (both for generic values of a braiding parameter $t$ and for the truncations occurring when $t$ coincides with certain roots of unity). The scheme corresponds to the canonical quantization of a set of fermionic oscillators which is supplemented by
the consistent information of their braiding properties.  The Hopf algebra under consideration is ${\cal U}({\mathfrak {gl}}(1|1))$, the Universal Enveloping Algebra of the ${\mathfrak{gl}}(1|1)$ superalgebra. The $t$-dependent braiding matrix $B_t$ entering the following construction is related, see \cite{kasa}, to the $R$-matrix of the Alexander-Conway polynomial. $B_t$ is recovered from the Burau representation \cite{bur} of the braid group (the connection between the Burau representation and the $R$-matrices of the quantum group ${\cal U}_q({\mathfrak {gl}(1|1))}$ has also been elucidated in \cite{rsw}). The special features of the representations of quantum groups at roots of unity are well known, see \cite{lus} and \cite{dck}.   In the present scheme the truncations at the roots of unity
are neatly derived from combinatorics which are easily proved via induction.  The roots of unity are obtained by solving an ordered set of polynomial equations (the solution of the $k$-th equation  implies that at most $k-1$ fermions can be accommodated in a multiparticle Hilbert space). \par
One should mention that the Hopf algebra scheme of  first quantization with braided tensor product was recently used to prove that the Rittenberg-Wyler \cite{{riwy1},{riwy2}} ${\mathbb Z}_2\times {\mathbb Z}_2$-graded  Lie (super)algebras lead to detectable parafermions  \cite{top1} and parabosons \cite{top2}. In those cases the braiding is simply given by sign assignments and not by a braiding matrix as here. \par
A summary of the results of the paper and its outlook will be given in the Conclusions. The possibility of a second quantization of braided Majorana fermions will be commented there.\par
Only the basic  properties of the braid group, Hopf algebras endowed with braided tensor product and Lie superalgebras are here recalled. Further  information is respectively found in \cite{bir}, \cite{maj} and  
\cite{kac}.\par
The scheme of the paper is the following. In Section {\bf 2} the Majorana fermions as ${\mathbb Z}_2$-graded qubits 
are introduced. The construction of the multiparticle Hilbert spaces is given in Section {\bf 3}. The truncations at the roots of unity are presented in Section {\bf 4} and the multiparticle Hilbert spaces and energy spectra in Section {\bf 5}. The future perspectives of the work are discussed in the Conclusions. 

\section{Majorana fermions as ${\mathbb Z}_2$-graded qubits}

Quantized Majorana fermions can be expressed, see \cite{kau}, via Clifford algebra generators. A ${\mathbb Z}_2$-graded qubit describes the Hilbert space  ${\cal H}^{(1)}$ of a single Majorana fermion. Let
$|vac\rangle$ be the bosonic vacuum and $|\psi\rangle$ the fermionic excited state; they are respectively given by 
{\footnotesize{\bea
|vac\rangle := \left(\begin{array}{c} 1\\0\end{array}\right) , &\quad& |\psi\rangle :=\left(\begin{array}{c} 0\\1\end{array}\right) .
\eea
}} 
The $2\times 2$ matrix operators acting on the graded qubit can be conveniently expressed in the following basis:
{\small{
\bea\label{abcd}
&\alpha =\left(\begin{array}{cc} 1&0\\0&0\end{array}\right),\quad ~\beta =\left(\begin{array}{cc} 0&1\\0&0\end{array}\right),\quad ~ \gamma =\left(\begin{array}{cc} 0&0\\1&0\end{array}\right),\quad ~\delta =\left(\begin{array}{cc} 0&0\\0&1\end{array}\right),&
\eea
}}
where $\alpha,\delta$ are even (bosonic) and $\beta,\gamma$ are odd (fermionic) matrices.
These four operators satisfy the following (anti)commutators which define the ${\mathfrak{gl}}(1|1)$ superalgebra:
\bea\label{anticomm}
&\relax [\alpha,\beta]=\beta, \qquad [\alpha,\gamma]=-\gamma,\qquad  [\alpha,\delta]=0,\qquad [\delta,\beta]=-\beta,\qquad  [\delta,\gamma]=\gamma,&\nonumber\\
&\{\beta,\beta\}=\{\gamma,\gamma\}=0,~~~\qquad\quad  \{\beta,\gamma\}=\alpha+\delta.&
\eea
The ${\mathbb Z}_2$-grading is given by
\bea
&{\mathfrak{gl}}(1|1)={\mathfrak{gl}}(1|1)_{[0]}\oplus{\mathfrak{gl}}(1|1)_{[1]}, \qquad {\textrm{with}} \quad \alpha,\delta\in{\mathfrak{gl}}(1|1)_{[0]}\quad {\textrm{and}}\quad \beta,\gamma\in {\mathfrak{gl}}(1|1)_{[1]}.&
\eea

In order not to burden the notation (the context will clarify which is which), the same set of $\alpha$, $\beta$, $\gamma$, $\delta$ symbols is used to denote the abstract ${\mathfrak{gl}}(1|1)$ generators satisfying the (\ref{anticomm}) (anti)commutators, as well as the $2\times 2$ matrices introduced in  (\ref{abcd}).\par 
The matrices $\gamma,\beta$ are a pair of fermionic creation/annihilation operators satisfying 
\bea
\{\gamma,\gamma\}=\{\beta,\beta\}=0, \quad \{\gamma,\beta\}= {\mathbb I}_2,  && \beta|vac\rangle=0,\quad
|\psi\rangle =\gamma|vac\rangle
\eea
(here and in the following ${\mathbb I}_n$ denotes the $n\times n$ Identity matrix).
  \par
Since bosons/fermions are superselected, the linear superposition of states belonging to different graded sectors is not allowed. Therefore,  the Hilbert space is graded and given by
\bea
{\cal H}^{(1)} = {\cal H}^{(1)}_{[0]}\oplus {\cal H}^{(1)}_{[1]}&\equiv& {\mathbb C}^{1|1}.
\eea
The elements of its even and odd sectors are respectively given by  
\bea
\qquad c_0|vac\rangle\in {\cal H}^{(1)}_{[0]}, \qquad c_1|\psi\rangle\in {\cal H}^{(1)}_{[1]}, &\qquad& {\textrm{with}} \qquad c_0,c_1\in {\mathbb C}.
\eea
A physical state is recovered by taking into account the irrelevance of the phase of a normalized vector. 
The above system describes two inequivalent physical states which are just $|vac\rangle$ and $|\psi\rangle$. They correspond to a classical
$1$ bit of information (off/on states).
Therefore, just like the physically inequivalent states of an ordinary qubit are specified by points of the ${\mathbf S}^2$ Bloch sphere, ${\mathbf Z}^2$ (which is equivalent to a classical bit) represents ``the Bloch sphere of the graded qubit".\par
By assuming the vacuum state to be bosonic and nondegenerate, the operator $\delta$ defined in (\ref{abcd}) can be regarded, without loss of generality, as the single-particle quantum Hamiltonian ${H}$:
\bea
{H}&:= &\gamma\beta=\delta= {\footnotesize{\left(\begin{array}{cc} 0&0\\0&1\end{array}\right)}}.
\eea
This choice of the Hamiltonian corresponds to set to $0$ the vacuum energy and to normalize to $1$ the energy eigenvalue
of the excited state.

\section{Construction of the braided multiparticle states}

We present the set of prescriptions which allow to construct the multiparticle Hilbert spaces of $N$, noninteracting, braided Majorana fermions within the Hopf algebra framework of \cite{maj}. \par
The ${\mathbb Z}_2$-graded $N$-particle Hilbert space ${\cal H}^{(N)}$ is a subset of the tensor product of $N$ single-particle Hilbert spaces ${\cal H}^{(1)}={\mathbb C}^{(1|1)}$. Let's set, for simplicity, ${\cal H}\equiv{\cal H}^{(1)}$. We have
\bea
{\cal H}^{(N)}&\subset &{\cal H}^{\otimes N}.
\eea
The $N$-particle vacuum state $|vac\rangle_{N}$ is the tensor product of $N$ single-particle vacua:
\bea
\qquad |vac\rangle_N &=& |vac\rangle \otimes \ldots \otimes |vac\rangle\qquad \qquad (\textrm{$N$ times}).
\eea
The construction of the multiparticle observables and excited states is made in terms of a special operation, the coproduct. Before going ahead, we briefly recall the main needed mathematical properties (more information can be found in \cite{maj} and also \cite{{top1},{top2}}).\par
In the present construction the  given Hopf algebra is a Universal Enveloping Algebra (denoted as  ${\cal U}\equiv {\cal U}({\mathfrak{g}})$) of a graded Lie algebra. A Hopf algebra is characterized by compatible structures (unit and multiplication), costructures (counit and coproduct) and antipode. Concerning the coproduct $\Delta$, it is a map
\bea\label{coproduct}
\Delta &:&{\cal U}\rightarrow {\cal U} \otimes  {\cal U}
\eea
which satisfies the coassociativity property
\bea\label{coassociativity}
   (\Delta\otimes id)\Delta(U)&=&(id\otimes \Delta)\Delta(U) \qquad {\textrm{for}}\quad U\in {\cal U},\nonumber\\
\Delta^{(n+1)} &=& (\Delta\otimes id)\Delta^{(n)}=(id\otimes \Delta)\Delta^{(n)}.
\eea
For any $U_A,U_B\in {\cal U}$, the further property
\bea\label{uaub}
   \Delta(U_AU_B)&=&\Delta(U_A)\Delta(U_B)
\eea
implies that the action on any given $U\in {\cal U}({\mathfrak g})$ is recovered from the action of the coproduct on the
Hopf algebra unit ${\bf 1}$ and the Lie algebra elements $g\in {\mathfrak{g}}$; they are given by
\bea\label{deltaidg}
   \Delta({\bf 1})={\bf 1}\otimes{\bf 1}, &\quad & 
   \Delta({ g})={\bf 1}\otimes{g}+g\otimes {\bf 1}.
\eea
Let $R$ be a representation of the Universal Enveloping Algebra ${\cal U}$ on a vector space $V$. The representation of the operators induced by the coproduct will be denoted with a hat:
\bea
&{\textrm{for}}\quad R: {\cal U}\rightarrow V,\qquad {\widehat \Delta}:= \Delta|_R\in End (V\otimes V),\qquad 
 {\textrm{with}}\quad  {\widehat{ \Delta(U)}}\in V\otimes V. &
\eea
It follows, from the second relation of (\ref{coassociativity}), that
\bea
{\widehat{ \Delta^{(n)}(U)}}&\in& V\otimes \ldots \otimes V\qquad (n+1 \quad {\textrm{times}}).
\eea
We are now in the position to define the multiparticle Hamiltonians.\par
The $N$-particle Hamiltonians $H_{(N)}$ are obtained by applying the $N$-particle coproducts $\Delta^{(N-1)}$ to the single-particle Hamiltonian $H=\delta$, while
an $N$-particle excited state is created by applying $\Delta^{(N-1)}$ to the creation operator $\gamma$. We have
\bea\label{ncoproduct}
H_{(N)} = {\widehat{\Delta^{(N-1)}(\delta)}}, && 
\gamma_{(N)} = {\widehat{\Delta^{(N-1)}(\gamma)}},
\eea
where in the above formulas the hat indicates that the coproduct is evaluated in the  representation (\ref{abcd}) of $\delta,\gamma$. Higher order excited states are recovered from the powers $\gamma_{(N)}^m$ with $m=1,2,\dots$. \par It turns out that
the $N$-particle observables and excited states are encoded in the ${\mathfrak{a}}\subset {\mathfrak{gl}}(1|1)$ ${\mathbb Z}_2$-graded subalgebra spanned by $\delta$ and $\gamma$:
\bea
\delta,\gamma\in {\mathfrak{a}}, && {\textrm{with brackets given by}} \quad [\delta,\gamma]=\gamma, \quad \{\gamma,\gamma\}=0.
\eea 
The ${\mathbb Z}_2$-graded Universal Enveloping Algebra under consideration is therefore ${\cal U}({\mathfrak a})$.\par
The (\ref{ncoproduct}) equations give, for $N=2,3,\ldots $, the formulas
\bea\label{h2and3}
H_{(2)} ={\mathbb I}_2\otimes\delta+\delta\otimes {\mathbb I}_2, \qquad\qquad~~&&~~\qquad\qquad \gamma_{(2)} ={\mathbb I}_2\otimes\gamma+\gamma\otimes {\mathbb I}_2,\nonumber\\
H_{(3)} ={\mathbb I}_2\otimes{\mathbb I}_2\otimes \delta+{\mathbb I}_2\otimes \delta\otimes {\mathbb I}_2+\delta\otimes{\mathbb I}_2\otimes {\mathbb I}_2, && \gamma_{(3)} ={\mathbb I}_2\otimes{\mathbb I}_2\otimes \gamma+{\mathbb I}_2\otimes \gamma\otimes {\mathbb I}_2+\gamma\otimes{\mathbb I}_2\otimes {\mathbb I}_2\nonumber\\&&
\eea
and so on.\par
The introduction of a non-trivial braiding requires specifying how Lie superalgebra generators are braided in a tensor product. The braiding of the elements of the Universal Enveloping Algebra  are obtained, see (\ref{uaub}), as a consequence. Let $a,b,c,d$ be four generators of a Lie superalgebra represented by $n$-dimensional matrices.  The construction of \cite{maj} can be expressed as
\bea
(a\otimes b)\cdot (c\otimes d) &=& (a\otimes {\mathbb I}_n)\cdot \Psi(b,c)\cdot ({\mathbb I}_n\otimes d),
\eea
where in the above formula $\Psi(b,c)$ is a $n^2\times n^2$ matrix which encodes the braiding of $b$ and $c$. The dots in the right hand side denote ordinary matrix multiplication. The consistency of the construction requires $\Psi(b,c)$ to satisfy the braiding conditions presented in \cite{maj}.\par
The construction of the multiparticle sectors and of the observables of the braided Majorana fermions only requires specifying the braidings of the two generators $\delta$ and $\gamma$. These generators can be identified with the $2\times 2$ matrices given in (\ref{abcd}).  The unique nontrivial braiding matrix is $\Psi(\gamma,\gamma)$; this is due to the fact that  it encodes the braiding properties of the Majorana fermions (we recall that $\gamma$ is their creation operator). We can therefore set
\bea\label{commuting}
&\Psi(\delta,\delta) = \delta\otimes \delta,\qquad \Psi(\delta,\gamma) =\gamma\otimes \delta,\qquad \Psi(\gamma,\delta) =\delta\otimes \gamma\quad &
\eea
and
\bea\label{ggbraiding}
\Psi(\gamma,\gamma) \equiv \Psi_t(\gamma,\gamma),~~ && {\textrm{ where, for ~}} t\in {{\mathbb C}^\ast}, \quad \Psi_t(\gamma,\gamma)=B_t\cdot \gamma\otimes \gamma .
\eea
$B_t$ is a $4\times 4$ constant matrix which depends on the parameter $t\neq 0$ and satisfies the braiding conditions; the dot in the right hand side of (\ref{ggbraiding}) denotes the standard matrix multiplication. In the above formula ${\mathbb C}^\ast\equiv {\mathbb C}\backslash\{0\}$ denotes the punctured complex plane without the origin.\par
A consistent choice for $B_t$ is
{\footnotesize{
\bea\label{bt}
B_t&=&\left(\begin{array}{cccc} 1&0&0&0\\0&1-t&t&0\\0&1&0&0\\0&0&0&-t\end{array}\right).
\eea
}}

As recalled in the Introduction, $B_t$ is related, see \cite{kasa} and \cite{rsw}, to both the Burau representation of the braid group and the $R$-matrix of the quantum group ${\cal U}_q({\mathfrak {gl}(1|1))}$.\par
The consistency of the (\ref{ggbraiding}) position for $\Psi_t(\gamma,\gamma)$ is guaranteed by the following braid relation being satisfied by $B_t$:
\bea
(B_t\otimes {\mathbb I}_2)\cdot ({\mathbb I}_2\otimes B_t)\cdot 
(B_t\otimes {\mathbb I}_2) &=& ({\mathbb I}_2\otimes B_t) \cdot
(B_t\otimes {\mathbb I}_2)\cdot ({\mathbb I}_2\otimes B_t).
\eea

The correctness of the (\ref{commuting}) positions is implied by $B_t$ being dynamically compatible, since it commutes with the $2$-particle Hamiltonian $H_{(2)}$ given in (\ref{h2and3}):
\bea
\relax [ H_{(2)}, B_t] &=& 0.
\eea
Formulas (\ref{commuting}) imply that, for any integer $N$, the $N$-particle creation operator $\gamma_{(N)}$ creates one quantum of energy:
\bea\label{nexcited}
\relax [H_{(N)}, \gamma_{(N)}] &=& \gamma_{(N)}.
\eea
Some comments are in order. The matrix $B_t$ is bosonic. Indeed, the even (odd) nonvanishing entries of the ${\mathfrak{gl}}(1|1)$ generators (\ref{abcd}) can be expressed as bullets (stars), so that in the tensor products we get
{\footnotesize{
\bea
\left(\begin{array}{cc} \bullet&\ast\\\ast&\bullet\end{array}\right)\otimes \left(\begin{array}{cc} \bullet&\ast\\\ast&\bullet\end{array}\right) &=& \left(\begin{array}{cccc} \bullet&\ast&\ast&\bullet\\\ast&\bullet&\bullet&\ast\\\ast&\bullet&\bullet&\ast\\\bullet&\ast&\ast&
\bullet\end{array}\right).\nonumber
\eea
}}

 $B_t$ is invertible for $t\neq 0$. The inverse is given by 
\bea
B_t^{-1} &=& B_1\cdot B_{\frac{1}{t}}\cdot B_1^{-1}.
\eea
At the special value $t=1$ one gets
\bea\label{tequal1}
&B_1=B_1^{-1}=B_1^T.&
\eea
The ordinary antisymmetry properties of the Majorana fermions are recovered, in the above construction, from 
the $t=1$ braiding matrix $\Psi_{t=1}(\gamma,\gamma)$.\par
The (unnormalized) vectors spanning the $N$-particle Hilbert spaces ${\cal H}^{(N)}_t$ of the braided Majorana fermions are given by
\bea\label{kvectors}
&|k\rangle _{t,{N}} =(\gamma)_{(N)}^k |vac\rangle_{N}\in {\cal H}^{(N)}_t, \qquad {\textrm{for}} \quad k=0,1,2,\ldots ,& 
\eea
so that $|0\rangle_{t,N}\equiv |vac\rangle_N$. The suffix $t$ denotes the choice of the braiding parameter entering (\ref{ggbraiding}); the braiding of the tensor products in the right hand side is given by $\Psi_t(\gamma,\gamma)$. In the following we will discuss under which conditions the vectors $|k\rangle _{t,{N}}$ are nonvanishing.\par
The energy eigenvalues $E_k$ are obtained from the equations
\bea\label{eneigen}
H_{(N)} |k\rangle _{t,{N}} &=& E_k |k\rangle _{t,{N}}.
\eea
From (\ref{ncoproduct}, \ref{nexcited}) we get, for a nonvanishing vector  $|k\rangle _{t,{N}}$:
\bea
E_k &=& k.
\eea
The energy eigenvalues, in particular, do not depend on the braiding parameter $t$.\par
Since $\gamma^2=0$, it easily follows for generic values of $t$ that $k=N$ produces the maximal energy eigenvalue of the $N$-particle states.
Indeed,  $(\gamma)_{(N)}^N\propto \gamma \otimes \gamma \otimes \ldots \otimes \gamma $, where the tensor product of $N$ $\gamma$'s is taken. For any integer $M>N$ we have $\gamma_{(N)}^M=0$.\par
The truncations of the energy spectrum for non-generic values of $t$ coinciding with roots of unity will be presented in the next Section.\par
The Fermion Number Operator $N_F$ admits $\pm 1$ eigenvalues. Bosonic (fermionic) states are defined to be the eigenvectors with $+1$ ($-1$) eigenvalue. For the models under consideration one can set
\bea\label{fernumberop}
N_F&=& (-1)^{H_{(N)}}.
\eea
It follows that the $|k\rangle _{t,{N}}$ states with even (odd) integer $k$ are bosons (fermions).

\section{Recursive relations and the truncations at roots of unity}

The $n$-th power of the $B_t$ matrix, for the integers $n\geq 1$, can be expressed as
{\small{
\bea
B_t^n&=&\left(\begin{array}{cccc} 1&0&0&0\\0&b_{n+1}(t)&t\cdot b_n(t)&0\\0& b_n(t)&t \cdot b_{n-1}(t)&0\\0&0&0&(-t)^n\end{array}\right),
\eea
}}
in terms of the functions $b_m(t)$, for $m=0,1,2,\ldots$, 
which satisfy the recursive relations
\bea\label{recursive}
&b_0(t)=0,\qquad b_1(t) =1, \qquad b_{n+1}(t) = (1-t)\cdot b_n(t)+t\cdot b_{n-1}(t).&
\eea 
It is easily checked by induction that $b_{n+1}(t)$ is given by
\bea\label{recursive2}
b_{n+1}(t) &=& \sum_{j=0}^{n} (-t)^j.
\eea
At the first orders we have $~b_2(t) =1-t,~$ $~b_3(t)=1-t+t^2~$ and so on.\par
We already remarked, see (\ref{tequal1}), that the $t=1$ point produces the $B_{t=1}$ square root of the identity matrix:
\bea
B_1^2 &=& {\mathbb I}_4.
\eea
The $n$-th roots of the identity matrix, so that 
\bea
B_t^n&=&{\mathbb I}_4,
\eea 
are encountered for the values of $t$ satisfying
\bea\label{roots}
&b_n(t)=0, \qquad{\textrm{together with}}\qquad b_{n+1}(t) =t\cdot b_{n-1}(t)= (-t)^n=1.&
\eea
The last equation implies, in particular, that $t$ should be  a root of unity.\par
Any choice of the $n-1$ roots of the polynomial equation $b_n(t)=0$ implies that the three remaining equations in (\ref{roots}) are automatically satisfied.\par
Indeed, we have that:\\
~
\\
{\it~~~ i}) $b_n(t)\equiv\sum_{j=0}^{n-1}(-t)^j=0$ implies that $(1+t)b_n(t)=0$ which, after straightforward manipulations,
leads to $(-t)^n=1$;\\
{\it~ ii}) $b_n(t)=0$ implies $tb_{n-1}(t)\equiv t\sum_{j=0}^{n-2} (-t)^{j}= 1 -\sum_{j=0}^{n-1} (-t)^j=1-b_n(t)=1$;\\
{\it iii}) $b_n(t)=0$, together with $tb_{n-1}(t)=1$, implies that the relation (\ref{recursive}) gives $b_{n+1}(t)=1$. \par
~\par
By expressing the roots of unity in terms of the angle $\vartheta\in [0, 2\pi[$ through the position $t=e^{i\vartheta}$,
we get that the $n-1$ solutions of $B_t^n={\mathbb I}_4$ are given, for the first few values of $n$, by
\bea\label{asets}
n=2 &:& \vartheta =0\in A,\nonumber\\
n=3 &:& \vartheta = \frac{1}{3}\pi, \frac{5}{3}\pi\in A,\nonumber\\
n=4 &:& \vartheta = \frac{1}{2}\pi, \frac{3}{2}\pi\in A \qquad {\textrm{and}}\quad  \vartheta = 0\in B, \nonumber\\
n=5 &:&\vartheta =\frac{1}{5}\pi, \frac{3}{5}\pi,\frac{7}{5}\pi, \frac{9}{5}\pi\in A,\nonumber\\
n=6 &:&\vartheta =\frac{2}{3}\pi,\frac{4}{3}\pi\in A\qquad {\textrm{and}}\quad \vartheta = 0, \frac{1}{3}\pi,\frac{5}{3}\pi\in B
\eea
and so on. For each $n$ we separated the roots into two sets, $A$ and $B$.
The roots belonging to the $A$ sets are ``new" roots first encountered at the value $n$, while the roots belonging to the $B$ sets are already found for some previous value $n'<n$ (e.g., $\vartheta = \frac{1}{3}\pi $ is encountered at $n=3 $ and $n=6$ and belongs, for the latter value, to the $B$ set).\par
Up to $n=9$, the total number of ``new" and remaining roots is given by the sums
\bea&
\begin{array}{ll} 
n=2:  \quad {\underline 1}+0,\qquad\quad &n=6:  \quad {\underline 2}+3,\\
n=3:  \quad {\underline 2}+0,\qquad\quad &n=7:  \quad {\underline 6}+0,\\
n=4:  \quad {\underline 2}+1,\qquad\quad &n=8:  \quad {\underline 4}+3,\\
n=5:  \quad {\underline 4}+0,\qquad\quad &n=9:  \quad {\underline 6}+2.\\
\end{array}&
\eea
In the above table the new roots belonging to the $A$ sets are underlined.\par
The $t$ roots of unity which solve the $B_t^n={\mathbb I}_4$ equation imply truncations of the multiparticle Hilbert spaces and energy spectra of the braided Majorana fermions. 

\subsection{The truncations at roots of unity}

The  $t$ roots of unity which satisfy the polynomial equations (\ref{roots}) produce truncations in multiparticle Hilbert spaces (and corresponding energy spectra) of the braided Majorana fermions. The truncations are obtained from the following combinatorics.\par
We recall at first that the $N$-particle Hilbert space is spanned by the vectors $|k\rangle _{t,{N}}$ given in (\ref{kvectors}); these vectors are obtained by applying to the $N$-particle vacuum $|vac\rangle_N$  the $k$-th powers $(\gamma_{(N)})^k$ of $\gamma_{(N)}$.  Concerning the braided tensor, it follows from (\ref{abcd}) and (\ref{bt}) that the braiding $\Psi_t(\gamma,\gamma) $,  given by (\ref{ggbraiding}), reads as
\bea
&({\mathbb I}_4\otimes\gamma)\cdot (\gamma\otimes{\mathbb I}_4) = \Psi_t(\gamma,\gamma) = -t\gamma\otimes\gamma.&
\eea
By taking into account that $\gamma^2=0$ simple computations show that, for $N=2,3$, the only nonvanishing powers of $\gamma_{(N)}$ are
\bea
\gamma_{(2)} &=&1\cdot({\mathbb I}_2\otimes \gamma+ \gamma\otimes {\mathbb I}_2),\nonumber\\
\gamma_{(2)}^2 &=& (1-t)\cdot( \gamma\otimes \gamma),\nonumber\\
\gamma_{(3)} &=&1\cdot({\mathbb I}_2\otimes{\mathbb I}_2\otimes \gamma+{\mathbb I}_2\otimes \gamma\otimes {\mathbb I}_2+\gamma\otimes{\mathbb I}_2\otimes {\mathbb I}_2),\nonumber\\
\gamma_{(3)}^2 &=& (1-t)\cdot ({\mathbb I}_2\otimes\gamma\otimes \gamma+\gamma\otimes{\mathbb I}_2\otimes\gamma+\gamma\otimes \gamma\otimes{\mathbb I}_2),\nonumber\\
\gamma_{(3)}^3 &=& (1-t)(1-t+t^2)\cdot (\gamma\otimes \gamma\otimes\gamma).
\eea
This result  is generalized by induction. Let us introduce the $A_{N;k}$ symbol which denotes the totally symmetrized sum of {\footnotesize{$\left(\begin{array}{c}N\\k\end{array}\right)$}} terms, the tensor products of $k$ matrices $\gamma$'s and $N-k$ identity matrices ${\mathbb I}_2$. The symbol is defined as
\bea
A_{N;k}&:=&\underbrace{ {\mathbb I_2}\otimes \ldots\otimes {\mathbb I}_2}_\text{$N-k$ times}\otimes \underbrace{\gamma\otimes \ldots \otimes \gamma}_\text{$k$ times} ~+~{\textrm{symmetrized terms}},
\eea
where $N$ is a positive integer, while $k$ is restricted to be $k=1,2,\ldots, N$. \par
At $k=1$, the $A_{N;1}$ symbol coincides with the $\gamma_{(N)}$ matrices introduced in (\ref{ncoproduct}):
\bea
A_{N;1} &=& \gamma_{(N)}.
\eea
It is proved by induction for $k\rightarrow k+1$ that any power $(\gamma_{(N)})^k$ is proportional to
$A_{N;k}$ with a normalizing factor $f_k(t)$ which does not depend on $N$. We have
\bea\label{fkrecursive}
(\gamma_{(N)})^k&=& f_k(t)\cdot A_{N;k}\qquad {\textrm{for}}\nonumber\\
f_k(t) &=& \prod_{j=1}^k b_j(t),
\eea
where the $b_j(t)$'s are the recursive polynomials (\ref{recursive},\ref{recursive2}). \par
At the lowest orders of $k=2,3,4,\ldots $ we have
\bea
(\gamma_{(N)})^2&=& (1-t)\cdot A_{N;2},\nonumber\\
(\gamma_{(N)})^3&=& (1-t)(1-t+t^2)\cdot A_{N;3},\nonumber\\
(\gamma_{(N)})^4&=& (1-t))(1-t+t^2)(1-t+t^2-t^3)\cdot A_{N;4}
\eea
and so on.\par
We limit here to sketch the general algorithmic proof of (\ref{fkrecursive}) which is based on braiding the $\gamma$'s entering the products $A_{N;1}\cdot A_{N;k}$. Let's exemplify the $N=5$, $k=2$ case. For simplicity we set $I\equiv {\mathbb I}_2$ and drop the unnecessary ``$\otimes$" symbol in the tensor product. We can then write
{\footnotesize\bea
A_{5;1}\cdot A_{5;2}&=&(\gamma IIII+I\gamma III+II\gamma II+III\gamma I+IIII\gamma)\cdot\nonumber\\
&&\cdot (\gamma \gamma III+\gamma I\gamma II+\gamma II \gamma I +\gamma III\gamma+ I\gamma\gamma II + I\gamma I\gamma I +I\gamma II\gamma +II\gamma\gamma I+ II\gamma I \gamma + III\gamma\gamma).\nonumber\\
&&
\eea
}} 
The result is $(1-t+t^2)\cdot A_{5;3} =b_3(t)\cdot A_{5;3}$. Indeed, by taking into account that $\gamma^2=0$, the contribution $\gamma\gamma\gamma II$ is obtained, e.g.,  by braiding the three tensor products $$\gamma IIII\cdot I\gamma\gamma II + I\gamma III\cdot \gamma I \gamma II+II\gamma II\cdot \gamma\gamma III,$$
so that the above expression can be written as $(1-t+t^2)\cdot \gamma\gamma\gamma II$.\par
~\par
The (unnormalized) multiparticle wavefunctions can be expressed as
\bea\label{vectorswithfa}
|k\rangle _{t,{N}}&=& f_k(t)\cdot A_{N;k}|vac\rangle_N.
\eea
The right hand side is nonvanishing for generic values of $t\neq 0$. Different Hilbert space truncations are obtained when $t$ coincides with a root of one of the recursive polynomials entering $f_k(t)$.

\section{The multiparticle Hilbert spaces and energy spectra}

The construction and combinatorics  presented in the two previous Sections bring, for multiparticle Hilbert spaces and energy spectra, the following results.\par
The roots of the polynomial equations $b_m(t)=0$, where the $b_m(t)$'s are the recursive polynomials (\ref{recursive},\ref{recursive2}), are organized into integer-labeled levels. We recall that, from the last equation of (\ref{roots}), any such root is necessarily a root of unity, while the converse is not true ($t=-1$, e.g.,  is not a solution of any $b_m(t)=0$ polynomial equation since, see formula (\ref{recursive2}), $b_n(-1)=n$; this implies that for any $n$, $B_{t=-1}^n \neq {\mathbb I}_4$). \par 
We define as ``level-$k$" root of unity, for $k=2,3,4,\ldots$, a solution $t_k$ of the $b_k(t_k)=0$ equation such that, for any  $k'<k$, $b_{k'}(t_k)\neq 0$. The first few level-$k$ roots of unity, up to $k=6$,  are given in (\ref{asets}); in that formula they correspond to the roots belonging to the $A$ sets. \par
From result ${\it i}$) of Section {\bf 4} follows that, when $k$ is a prime number, the $k-1$ solutions of the $b_k(t)=0$ equation are all level-$k$ roots of unity.\par
The physical significance of a level-$k$ root of unity lies in the fact that the corresponding braided multiparticle Hilbert space can accommodate at most $k-1$ Majorana spinors. This is read from the vanishing condition for $|k\rangle _{t,{N}}$ in equation (\ref{vectorswithfa}) which takes into account that the
proportionality factor $ f_k(t)$ in the right hand side is the product $f_k(t) = \prod_{j=1}^k b_j(t)$.\par
 The special point $t=1$, being the solution of the $b_2(t)\equiv 1 -t=0$ equation, is a level-$2$ root of unity. It corresponds to the ordinary total antisymmetrization of the fermionic wavefunctions.  In this framework the $t=1$ level-$2$ root of unity encodes the Pauli exclusion principle of ordinary fermions.\par
We are now in the position to present the energy spectra of the multiparticle Hilbert spaces at varying $t\in {\mathbb C}^\ast$. They are computed from the (\ref{eneigen}) equation. It follows from  (\ref{vectorswithfa})
that the energy eigenvalues are not degenerate.\par
Before presenting the general result we illustrate some examples by giving some tables of the $N$-particle energy eigenvalues for the first few truncated level-$k$ roots of unity, up to
$k=5$.\par
 In the tables below the corresponding eigenvalues are marked with an ``$X$". The rows denote the energy levels $E$ and the columns the $N$-particle (up to $N=7$) Hilbert spaces. We have\par
~\par
{\it i}) level $k=2$ root of unity: the unique case is  $t=1$; the $N$-particle energy levels are
\bea\label{level2}
&\begin{array}{|c|c|c|c|c|c|c|c|}\hline E\backslash N&1&2&3&4&5&6&7\\ \hline  2&&&&&&&\\ \hline 1&X&X&X&X&X&X&X\\ \hline 0&X&X&X&X&X&X&X\\ \hline
\end{array}&
\eea
{\it Comment}: this table corresponds to the ordinary, totally antisymmetrized, Majorana fermions, with only  $E=0,1$ energy eigenvalues for any $N$.\par
~\par
{\it ii}) level $k=3$ roots of unity, given by $t=e^{i\vartheta}$ with $\vartheta =\frac{1}{3}\pi,\frac{5}{3}\pi$; the energy eigenvalues are
\bea\label{level3}
&\begin{array}{|c|c|c|c|c|c|c|c|}\hline E\backslash N&1&2&3&4&5&6&7\\ \hline 3&&&&&&&\\ \hline 2&&X&X&X&X&X&X\\ \hline 1&X&X&X&X&X&X&X\\ \hline 0&X&X&X&X&X&X&X\\ \hline
\end{array}&
\eea
{\it Comment}: the energy eigenvalues are $E=0,1,2$ for any multiparticle sector with $N\geq 2$.\par
~\par
{\it iii}) level $k=4$ roots of unity, given by  $t=e^{i\vartheta}$ with $\vartheta =\frac{1}{2}\pi,\frac{3}{2}\pi$; the energy eigenvalues are
\bea\label{level4}
&\begin{array}{|c|c|c|c|c|c|c|c|}\hline E\backslash N&1&2&3&4&5&6&7\\ \hline \hline 4&&&&&&&\\ \hline 3&&&X&X&X&X&X\\ \hline 2&&X&X&X&X&X&X\\ \hline 1&X&X&X&X&X&X&X\\ \hline 0&X&X&X&X&X&X&X\\ \hline
\end{array}&
\eea
{\it Comment}: a ``plateau" is reached; starting from $N\geq 3$ the energy eigenvalues are $E=0,1,2,3$.
\par
~\par
{\it iv}) level $k=5$ roots of unity given by $t=e^{i\vartheta}$, $\vartheta =\frac{1}{5}\pi,\frac{3}{5}\pi,\frac{7}{5}\pi,\frac{9}{5}\pi$; the energy eigenvalues are
\bea\label{level5}
&\begin{array}{|c|c|c|c|c|c|c|c|}\hline E\backslash N&1&2&3&4&5&6&7\\ \hline 5&&&&&&&\\ \hline
4&&&&X&X&X&X\\ \hline 3&&&X&X&X&X&X\\ \hline 2&&X&X&X&X&X&X\\ \hline 1&X&X&X&X&X&X&X\\ \hline 0&X&X&X&X&X&X&X\\ \hline
\end{array}&
\eea

{\it Comment}: the plateau is shifted at $N\geq 4$, with energy eigenvalues $E=0,1,2,3,4$.\par
~\par

We present now the general formulas which describe two different physical regimes and are therefore divided into two distinct subcases.\par
~\par
{\bf Subcase {\it a}}, $N$-particle energy eigenvalues $E$ for $t$ belonging to a level-$k$ root of unity:
\bea\label{truncatedenergy}
E &=& 0,1,\ldots, N\qquad \quad~{\textrm{for}}\quad N<k,\nonumber\\
E &=& 0,1,\ldots, k-1 \qquad {\textrm{for}} \quad N\geq k.
\eea
{\it Comment}: the plateau is reached for the maximal energy level $k-1$; this is the maximal number of braided Majorana fermions that can be accommodated in a multiparticle Hilbert space.\par
~\par
{\bf Subcase {\it b}}, $N$-particle energy eigenvalues $E$ for a generic value $t\in {\mathbb C}^\ast$ which does not coincide with a level-$k$ root of unity:
\bea\label{genericenergy}
E &=& 0,1,\ldots, N\qquad {\textrm{for any}}\quad N.
\eea

{\it Comment}: in this subcase there is no plateau; the energy eigenvalues grow linearly with $N$.

\subsection{Generalized Bloch spheres and space of states}

The Hilbert spaces are ${\mathbb Z}_2$-graded and split into even (bosonic) and odd (fermionic) sectors, defined by the Fermion Number Operator $N_F$ introduced in (\ref{fernumberop}). The projectors $P_\pm$, given by
\bea
P_\pm &=& \frac{1}{2}({\mathbb I}\pm N_F),\qquad {\textrm{so that}}\quad P_++P_-={\mathbb I},\quad P_\pm^2=P_\pm \quad {\textrm{and}}\quad P_+P_-=P_-P_+=0,\nonumber\\&&
\eea
allow to define a superselection rule; the bosonic $|bos\rangle$ and fermionic $|fer\rangle$ states  are defined to respectively satisfy
\bea
P_+|bos\rangle =|bos\rangle, \quad &&\quad P_-|fer\rangle = |fer\rangle.
\eea
The superselection means that one cannot make a linear combination of bosonic and fermionic states. Only states in the given sector (bosonic or fermionic) can be linearly combined.\par
The energy levels of the multiparticle sectors are presented in formulas (\ref{truncatedenergy},\ref{genericenergy}). The energy levels $E$'s are given by the non-negative integers 
 $E=0,1,\ldots, s$ up to a maximal value $s$; the value $s=1,2,3,\ldots$ is determined by both the braiding parameter $t$ and the total number $N$ of particles in the multiparticle sector. Accordingly, the graded Hilbert spaces are respectively given, for odd and even values of $s$, by
\bea
{\mathbb C}^{r|r}\quad{\textrm{for}}\quad s+1=2r, &\quad&
{\mathbb C}^{r+1|r}\quad{\textrm{for}}\quad s=2r.
\eea
It follows that the Hilbert subspaces of each superselected (bosonic or fermionic) graded sector are identified with either ${\mathbb C}^r$ or (for the bosonic sector with $s=2r$) ${\mathbb C}^{r+1}$; we obtain  ${\mathbb C}^{1|1}$ for $s=1$, ${\mathbb C}^{2|1}$ for $s=2$, ${\mathbb C}^{2|2}$ for $s=3$ and so on.\par
As discussed in Section {\bf 2}, generalized Bloch spheres define the space of physical states recovered from the normalization of the vectors and the irrelevance of the phases. By taking into account the superselection of the graded sectors, the inequivalent physical states are expressed by the pairs $(x,0)$ for the bosonic sector  and $(0,y)$ for the fermionic sector, where the points $x,y$ are constrained. At the lowest values of $s$ the constraints are:
\begin{itemize}
\item for $s=2$ the pairs are $(1,0)$ and $(0,1)$,
\item for $s=3$ the pairs are $(x\in {\bf S}^2,0)$ and $(0,1)$,
\item  for $s=4$ the pairs are $(x\in {\bf S}^2,0)$ and $(0,y\in{\bf S}^2)$,
\item for $s=5$ the pairs are $(x\in {\bf \Omega}_3,0)$ and $(0,y\in{\bf S}^2)$.
\end{itemize}
The extension to higher values $s>5$ is immediate.\par
At $s=5$ the inequivalent bosonic physical states are identified with the points of the ${\bf \Omega}_3$ set
which is, see \cite{gnss}, the ``generalized Bloch sphere of a qutrit".

\section{Conclusions}

The paper presents, within the \cite{maj} framework of graded Hopf algebras endowed with a braided tensor product, the First Quantization of braided Majorana fermions.\par
The relevant braiding matrix $B_t$ is related, see \cite{{kasa},{rsw}}, to both the Burau representation of the braid group and the $R$-matrix of the quantum group ${\cal U}_q({{\mathfrak gl}}(1|1))$. Two different physical regimes appear.  In the first one, truncations of the multiparticle Hilbert spaces are encountered when the braiding parameter $t$ is a root of unity which solves one of the
polynomial equations of the recursive set (\ref{recursive},\ref{recursive2}). These roots of unity are organized into levels which specify, see (\ref{truncatedenergy}),
the maximal number of allowed braided Majorana fermions in a multiparticle sector. The second regime, see formula (\ref{genericenergy}), is for generic values of the braiding parameter $t\in {\mathbb C}^\ast$.\par
The Hilbert spaces are ${\mathbb Z}_2$-graded and superselected. At $t=1$ the ordinary totally antisymmetric wavefunctions, which imply the Pauli exclusion principle for fermions, are recovered. A nontrivial $t\neq 1$ braiding is required, see subsection ({\bf 5.1}), in order to accommodate in the multiparticle sectors qubits, qutrits and so on.\par
Some comments are in order. The truncated models under consideration can also be seen as a specific implementation of the parafermionic statistics, see \cite{{gre},{grme}}, which allows to accommodate at most a given finite number of parafermions in any multiparticle sector. The parastatistics framework of \cite{{gre},{grme}} is based on the trilinear relations. The connection between the Hopf algebra's approach to parastatistics and trilinear relations has been discussed in \cite{{anpo1},{kada1}}.\par
It should be pointed out that the present first quantization framework can be extended to more general representations of the braid group. It was shown in \cite{coto} that the Artin's faithful braid representation as automorphisms on a free group \cite{art} can be linearized by applying the Magnus representation of a free group \cite{mag}. This construction leads to the so-called ``braid lift" presented in \cite{luto}: an infinite tower of braid representations
and Yang-Baxter matrices are iteratively produced from lower order representations. Starting from the trivial representation of the braid group, the first lift produces the Burau representation and its associated Yang-Baxter matrix, while the second lift gives the Lawrence representation of Hecke type \cite{law}.\par
A further motivation of this work is the prospect of establishing a connection with a possible Second Quantization of braided Majorana fermions. Following an argument discussed in  \cite{tod}, the Hopf algebra framework based on coproduct cannot include particle interactions. Quite likely, this argument can be circumvented for integrable systems (since they can be linearized with the appropriate choice of action-angle coordinates) but conserves its general validity; it then suggests that a full Second Quantization should be pursued. Quite recently a framework for noncommutative braided field theories, based on $L_\infty$-algebras, was made available (see the review\cite{gisz} and the references therein). It looks promising 
to exploit it for a Second Quantization. 
Working out the details of a First Quantization, as done here,  is the necessary propaedeutic step towards
this research program.

\par
~\par
~
\\ {\Large{\bf Acknowledgments}}
{}~\par{}~\\
I am grateful to Zhanna Kuznetsova for discussions and suggestions.\\
 This work was supported by CNPq (PQ grant 308846/2021-4).


\begin{thebibliography}{99}


\bibitem{kit} A. Yu. Kitaev, {\it Fault-tolerant quantum computation by anyons}, Ann. of Phys. {\bf 303},  2 (2003);
arXiv:quant-ph/9707021.
\bibitem{brki} S. B. Bravyi and A. Yu. Kitaev, {\it Fermionic quantum computation}, Ann. of Phys.  {\bf 298}, 210 (2002); arXiv:quant-ph/0003137.
\bibitem{kau} L. H. Kauffman, {\it Knot logic and topological quantum computing with Majorana fermions},  in 
``{\it Logic and Algebraic Structures in Quantum Computing}", p. 223, Cambridge Univ. Press (2016); arXiv:1301.6214[quant-ph].
\bibitem{maj}
S. Majid, {\it Foundations of Quantum Group Theory}, Cambridge University Press, Cambridge (1995).
\bibitem{kalo} L. H. Kauffman and S. J. Lomonaco, {\it Braiding, Majorana fermions, Fibonacci particles and topological quantum computing}, Quantum. Inf. Process. {\bf 17}, 201 (2018).
\bibitem{gyxz} M. L. Ge, L. W. Yu, K. Xue and Q. Zhao, {\it Yang-Baxter equation, Majorana fermions and three body 
entangling states}, Int. J. Mod. Phys. {\bf B 28}, 1450089 (2014). 
\bibitem{geli} L. W. Yu and M. L. Ge, {\it More about the doubling degeneracy operators associated with
Majorana fermions and Yang-Baxter equation}, Sci. Rep. {\bf 5}, 8102 (2015); arXiv:1409.3396[quant-ph].
\bibitem{kasa} L. Kauffman and H. Saleur, {\it Free fermions and the Alexander-Conway polynomial}, Comm. Math. Phys. {\bf 141}, 293 (1991).
\bibitem{bur} W. Burau,  {\it \"Uber Zopfgruppen und gleichsinnig verdrillte Verkettungen}, Abh. Math. Semin. Univ. Hambg. {\bf 11}, 179 (1935).
\bibitem{rsw} N. Reshetikhin, C. Stroppel and B. Webster, {\it Schur-Weyl-Type Duality for Quantized $gl(1|1)$, the Burau Representation of Braid Groups, and 
Invariant of Tangled Graphs}, in ``{\it Perspective in Analysis, Geometry, and Topology}" (PM, Vol. {\bf 296}), Birkh\"auser, p.  389 (2012); arXiv:1903.03681[math.RT].
\bibitem{lus} G. Lusztig, {\it Quantum groups at roots of $1$}, Geom. Dedicata {\bf 35}, 89 (1990).
\bibitem{dck} C. de Concini and V. G. Kac, {\it Representations of quantum groups at roots of $1$}, in ``{\it Operator
Algebras, Unitary Representations, Enveloping Algebras, and Invariant Theory}", ed. A. Connes et al., Birkh\"auser, p. 471 (2000).

\bibitem{riwy1} V. Rittenberg and D. Wyler, 
{\it Generalized Superalgebras}, 
{Nucl. Phys.} {\bf B 139}, 189 (1978).
\bibitem{riwy2} V. Rittenberg and D. Wyler, 
{\it Sequences of $Z_2\otimes Z_2$ graded Lie algebras and superalgebras}, 
{J. Math. Phys.} {\bf 19}, 2193 (1978).

\bibitem{top1} F. Toppan, {\it 
\zzg parastatistics in multiparticle quantum Hamiltonians}, J. Phys. A: Math. Theor. {\bf 54}, 115203 (2021);
arXiv:2008.11554[hep-th].

\bibitem{top2} F. Toppan, {\it Inequivalent quantizations from gradings and
\zzg parabosons}, J. Phys. A: Math. Theor.  {\bf 54}, 355202 (2021);
arXiv:2104.09692[hep-th].

\bibitem{bir} J. S. Birman, {\it Braids, Links and Mapping Class Groups}, (AM-82) Vol. {\bf 82}, Princeton Univ. Press, Princeton (NJ)  (1975).
\bibitem{kac}  V. G. Kac, {\it Lie Superalgebras}, Adv. in Math. {\bf 26}, 8 (1977).


\bibitem{gnss} S. K. Goyal, B. Neethi Simon, R. Singh and S. Simon, {\it Geometry of the generalized Bloch sphere for qutrits}, J. Phys. A: Math. Theor. {\bf 49}, 165203 (2016); arXiv:1111.4427[quant-ph].


\bibitem{gre} H. S. Green, ``{\it A Generalized Method of Field Quantization}", Phys. Rev. {\bf 90}, 270 (1953).
\bibitem{grme} O. W. Greenberg and A. M. L. Messiah, ``{\it Selection Rules for Parafields and the Absence of Para Particles in Nature}", Phys. Rev. {\bf 138}, B 1155 (1965).


\bibitem{anpo1} B. Aneva and T. Popov, {\it Hopf Structure and Green Ansatz of Deformed Parastatistics Algebras},
J. Phys {\bf A}: Math. Gen. {\bf 38}, 6473 (2005); arXiv:math-ph/0412016.
\bibitem{kada1} K. Kanakoglou and C. Daskaloyannis, {\it Parabosons quotients. A braided look at Green's ansatz and a generalization}, J. Math. Phys. {\bf  48}, 113516 (2007); arXiv:0901.04320[math-ph].

\bibitem{coto} F. Constantinescu and F. Toppan, {\it On the Linearized Artin Braid Representation}, J. Knot Th. and Its Ramifications {\bf 2} n. {4}, 399 (1993); arXiv:hep-th/9210020.


\bibitem{art} E. Artin, {\it Theorie der Z\"opfe}, Abh. Math. Semin. Univ. Hambg. {\bf 4}, 47 (1925).


\bibitem{mag} W. Magnus, {\it On a Theorem of Marshall Hall},  Ann. of Math. {\bf 40}, 764 (1939).



\bibitem{luto} M. L\"udde and F. Toppan, {\it Matrix solutions of Artin's braid relations}, Phys. Lett. {\bf B 288}, 321 (1992).

\bibitem{law} R. J. Lawrence, {\it Homological representations of the Hecke algebra}, Comm. Math. Phys. {\bf 135},
141 (1990).
\bibitem{tod} I. Todorov, {\it Quantization is a mystery}, Bulg. J. Phys. {\bf 39}, 107 (2012); arXiv:1206:3116[math-ph].
\bibitem{gisz} G. Giotopoulos and R. J. Szabo, {\it Braided Symmetries in Noncommutative Field Theory}, arXiv:2112.00541[hep-th].

\end{thebibliography}
\end{document}